\documentclass{svjour3}                     

\smartqed  
\usepackage{graphicx}
 \journalname{J. Low. Temp. Phys.}

\newcommand{\rem}[1]{}
\newcommand{\refe}[1]{(\ref{#1})}

\newcommand{\refE}[1]{Eq.~(\ref{#1})}
\newcommand{\beq}{\begin{equation}}
\newcommand{\eeq}{\end{equation}}
\newcommand{\beqa}{\begin{eqnarray}}
\newcommand{\eeqa}{\end{eqnarray}}

\begin{document}

\title{Cooling a vibrational mode coupled to a molecular single-electron transistor}

\author{F. Pistolesi}


\institute{F. Pistolesi \at
    Laboratoire de Physique et Mod\'elisation des Milieux
    Condens\'es,\\
    CNRS and Université Joseph Fourier
    B.P. 166, F-38042 Grenoble, France \\
    \email{fabio.pistolesi@grenoble.cnrs.fr}           
}

\date{Received: date / Accepted: date}

\maketitle

\begin{abstract}
We consider a molecular single electron transistor coupled
to a vibrational mode.
For some values of the bias and gate voltage
transport is possible only by absorption of one
ore more phonons.
The system acts then as a cooler for the
mechanical mode at the condition that the
electron temperature is lower than the
phonon temperature.
The final effective temperature of the vibrational mode
depends strongly on the bias conditions and
can be lower or higher of the reservoir in contact
with the oscillator.
We discuss the efficiency of this method, in particular we
find that there is an optimal value for the electron-phonon
coupling that maximizes cooling.
\end{abstract}

\section{Introduction}

Cooling the vibrational mode of
nano-mechanical oscillator is a major challenge on the road
to the observation of quantum effects in mechanical systems.
Recently several proposals for cooling mechanical modes have been
formulated
\cite{martin:2004,wilson-rae:2004,clerk:2005,blencowe:2005,wilson-rae:2007,marquardt:2007,rajauria:2007,rodrigues:2007,genes:2008,dantan:2008,metzger:2008,hekking:2008,hauss:2008,ying-dan:2008,koerting:2008,zippilli:2008}
and many have been realized
experimentally \cite{naik:2006,schliesser:2006,arcizet:2006,vinante:2008,teufel:2008}.
A particularly efficient method is the so called sideband cooling (see for instance
Refs. \cite{wilson-rae:2007,marquardt:2007}).
The general idea of side-band cooling is to exploit a detuning
between the frequency of a forcing field (for instance
a laser source) and the resonant frequency of a cavity mode.
The cavity mode is coupled to a mechanical oscillator of
resonant frequency $\omega$.
When the detuning is negative (the source-field frequency is lower than
the cavity mode frequency) and close to $\omega$ the increase of the
occupation of the cavity mode can be realized by the absorption
of a phonon of the oscillator.
This process removes energy from the mechanical mode and thus
cools the oscillator.
For positive detuning one observes the opposite
phenomenon of heating due to the
coherent pumping of energy in the mechanical mode.

\begin{figure}
  \includegraphics[width=6cm,angle=-90]{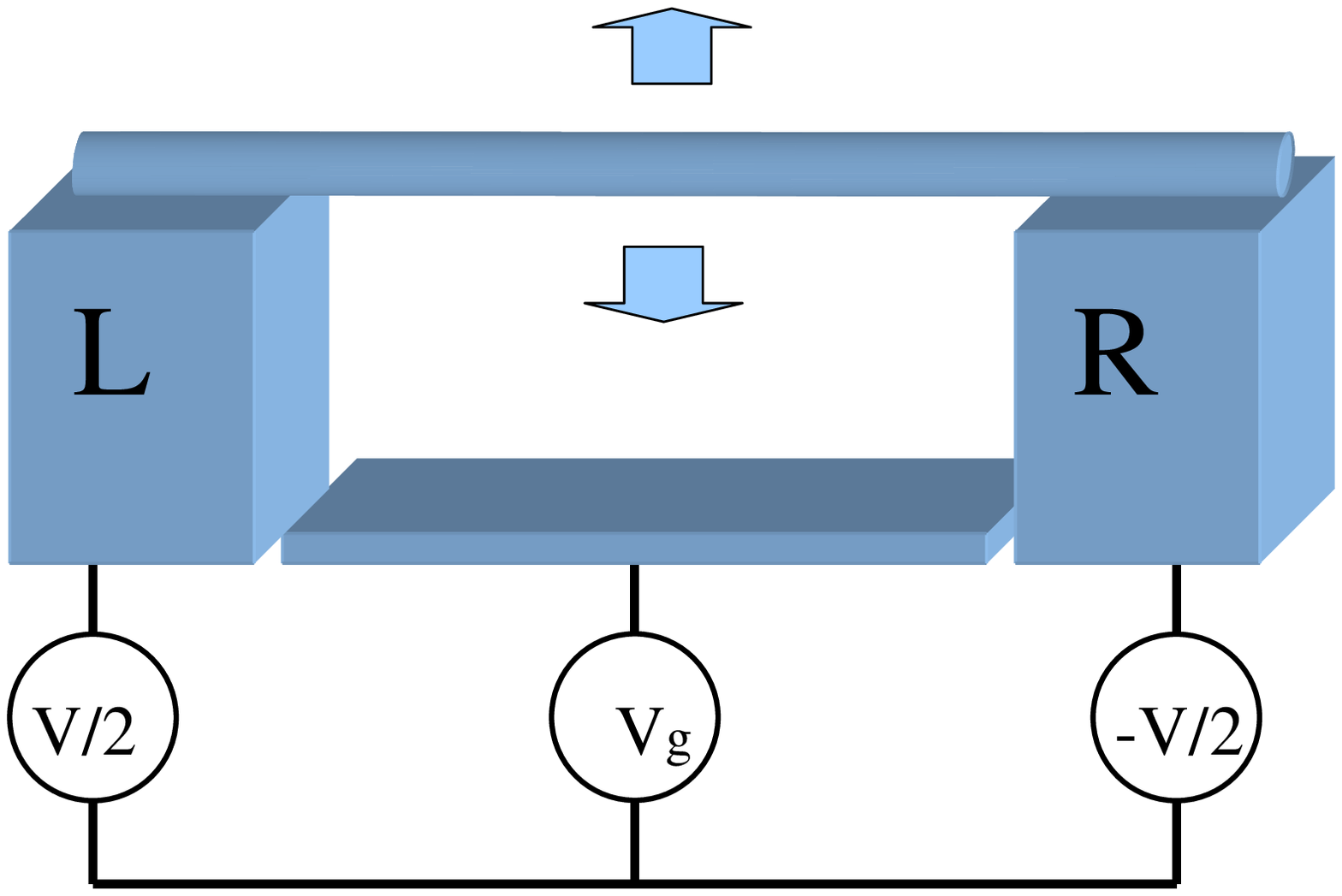}
\caption{An example of the system considered in the text,
a suspended nanotube contacted to two leads and coupled capacitively to a gate.
}
\label{device}       
\end{figure}

In this paper we consider a similar phenomenon that can take place
in a typical nano-electromechanical device: a molecular single
electron transistor coupled to a mechanical oscillator (see Fig. \ref{device}).
%
%
%
%
This system has been widely studied in the literature
\cite{park:2000,boese:2001,nitzan:2003,braig:2003,leroy:2004,mitra:2004,blanter:2004E,koch:2005,mozyrsky:2006,zazunov:2006}.
In particular it has been shown that the coupling to the mechanical
mode can induce a suppression of the current at low bias both in
the molecular
case \cite{braig:2003,koch:2005,mozyrsky:2006}
(when the size of the central island is so small that a
single electronic level participates in the transport)
and in the metallic classical case
(when the level spacing in the island is negligible with
respect to the bias voltage) \cite{pistolesi:2007}.
The suppression of the current is known as Frank-Condon blockade
or phonon-blockade, and it has a very simple classical origin:
When the island is charged the force acting on it
can displace the island itself leading to a different gate voltage ($V_g$)
seen by the device.
This corresponds to a shift of the Coulomb diamonds in the
$V_g$-$V$ plane ($V$ being the bias voltage) by a small amount in the $V_g$ direction.
The current can then flow only in the regions corresponding to the
intersection of the sequential transport regions of the shifted and
un-shifted diamonds.
By a simple geometrical construction one finds that
at low voltage current is suppressed \cite{pistolesi:2007}.
In the quantum limit $\hbar \omega \sim  k_B T, eV$, (where $T$ is the
environment temperature, $k_B$ is the Boltzmann constant
and $e$ is the electron charge) the shift of the diamonds is discrete
and each step corresponds to the absorption or the emission of phonons.
The resulting Coulomb diamonds are characterized by emission
\cite{braig:2003,koch:2005} or absorption sidebands \cite{luffe:2008}.
Very recently the emission sidebands predicted by these theories
have been observed in suspended carbon nanotubes \cite{leturcq:2008}.

In this paper we discuss under which conditions it is possible
to exploit the phonon-assisted tunneling to cool the oscillator.
A necessary condition is that the electronic temperature is lower than
the temperature of the environment coupled to the oscillator.
This fact can be realized either because the oscillator is coupled better
than the electrons to a hot environment, or by a preliminary cooling of the
electrons by existing techniques of electrons cooling \cite{leivo:1996,rajauria:2007}.
In this paper we explain how one could sequentially cool also
the mechanical mode by exploiting the properties of the single electron
transistor to optimize cooling.
We find that it exist regions in the $V_g$-$V$ plane
where the cooling effect is maximum, these are the regions just
outside the conducting diamonds within a distance in energy of the
order of $\hbar \omega$.
We also find that in order to achieve a good cooling efficiency the electron-phonon
coupling constant must not be too large, since then the same mechanism at the
base of the phonon blockade blocks transport at low bias thus reducing also the
number of absorbed phonons.

The paper is organised as follows: In Section \ref{sec2} we describe
in details the system and the theoretical approach in terms of a master equation
for the phonon distribution.
In Section \ref{sec3} we give an analytical description of cooling for small bias
voltage and near the border of the conducting region in the $V_g$-$V$ plane.
In Section \ref{sec4} we discuss the numerical solution of
the master equation for several values of the parameters.
Section \ref{sec5} gives our conclusions.

\section{System model}
\label{sec2}

We consider large molecules (like carbon
nanotubes) or quantum dots connected to two leads and in presence of a gate
voltage (see Fig. \ref{device}).
If the Coulomb charging energy is sufficiently high this system acts as a single
electron transistor, where transport can be finely tuned by varying the gate
voltage.
We assume that the presence of an excess electron in the central island generates
a force on a nano-oscillator, that can be a vibrational mode of the molecule
itself or a nearby mobile and gated lead.
We consider the regime where the oscillation frequency is larger or of the same
order of the temperature, the bias and the gate voltage.
This regime can be easily realized for small molecules, where the oscillating
frequency can attain the THz \cite{park:2000,leroy:2004},
in larger molecules and for metallic oscillators this can be more difficult since
for instance a long carbon nanotube the oscillation frequency can be in the MHz range
\cite{sazanova:2004,sapmaz:2006} rendering the system classical \cite{pistolesi:2007}.
A minimal model for such a system is the Anderson-Holdstein Hamiltonian
that has been considered by many
authors \cite{glazman:1988,wingreen:1989,braig:2003,mitra:2004,zazunov:2007}:
\beq
    H = [\epsilon_0 + \lambda \hbar \omega(\hat{a}+\hat{a}^\dag)] \hat{d}^\dag \hat{d}
    + \hbar \omega \hat{a}^\dag \hat{a}
    + \sum_{k,\alpha}{\epsilon_{k\alpha} \hat{c}^\dag_{k\alpha}\hat{c}^{\vphantom{\dag}}_{k\alpha}} +
    \sum_{k,\alpha}{t_{\alpha} (\hat{c}^\dag_{k\alpha}\hat{d} +
    \hat{d}^\dag \hat{c}^{\vphantom{\dag}}_{k\alpha})}
    \,.
\eeq
Here $\alpha$ is the lead index ($L$ or $R$), $\hat{c}$, $\hat{c}^\dag$,
$\hat{d}$, and  $\hat{d}^\dag$ are the electron
annihilation and creation operators for the leads ($c$) and local orbital ($d$).
$\epsilon_{k\alpha}=\xi_k-\mu_\alpha$ is the dispersion relation
of the single-electron bands in the lead $\alpha$ with $\mu_\alpha$
being the chemical potential.
The index $k$ labels the single electron states in the two leads.
$\epsilon_0$ is the energy eigenvalue of the local orbital and can be
shifted by the external gate voltage (cfr. Fig. \ref{device}).
The operators $\hat{a}$ and $\hat{a}^\dag$ annihilate and create
phonons, $\lambda$ is the electron-phonon dimensionless coupling.
We consider the model for spinless electrons for simplicity.
Inclusion of spin along with large onsite Coulomb repulsion
leads to qualitatively similar results, as far as the Kondo temperature
$T_K \ll \hbar \omega, k_B T_e$, with $T_e$ the electronic temperature.
We also assume a wide band limit, so that the tunneling induced level
width is simply $\Gamma_\alpha = \pi \nu_\alpha t_\alpha^2$,
where $\nu_\alpha$ is the density of states and
$t_\alpha$ is the tunneling amplitude in lead $\alpha$.
The exact definition of the electron phonon coupling $\lambda$ depends on the
explicit geometry and material parameters of the device.
Values larger than unit are possible, they correspond to a displacement of the
mechanical oscillator over a distance larger than the zero point motion length
when an electron is added in the dot.
Very recently  $\lambda$ has been evaluated theoretically
for the elongation modes of a suspended nanotube \cite{leturcq:2008}.
In good agreement with the observed magnitude of the Frank-Condon sideband
the authors found that $\lambda\approx 1.3$, thus exactly in the intermediate range
relevant for the present paper.
The mechanical state for $k_B T_e, \Gamma_\alpha \gg \hbar \omega$ has been
discussed in Ref. \cite{mozyrsky:2006,pistolesi:2008} for the molecular case
and in Ref. \cite{armour:2004,doiron:2006,pistolesi:2007,usmani:2007} for the metallic case.
(The case of a single tunnel junction is discussed in Ref. \cite{pauget:2008}.)
The frequent hopping of electrons in and out the central
island are a strong source of decoherence for
the oscillator and the reduced density matrix is diagonal in
the position and momentum basis of the oscillator itself.
A classical approach is thus appropriate in this case.
Here we consider the opposite limit of fast phonons:
$\hbar \omega \gg {\rm min}(\Gamma_L,\Gamma_R)$
\cite{braig:2003,mitra:2004,koch:2005,ryndyk:2006}.
The mechanical oscillator on average performs
many revolutions between two electron hopping
events.
The oscillator reduced density matrix in first approximation is
then diagonal in the base of the energy eigenvectors (phonons).
The time evolution of the system can then be described by a
(Pauli) master equation for the probability $P_{np}$
that the central island is occupied by $n$ electrons (with
$n=0$ or $1$ in our case)
and the oscillator is in the eigenstate of the free oscillator Hamiltonian
with eigenvalue $n \hbar \omega$.
The rates from one state to the other can be calculated
with Fermi golden rule in second order in the tunneling
coupling and all orders in the interaction constant $\lambda$.
The standard approach is to perform a Lang-Firsov canonical transformation
to eliminate the coupling term proportional to $\lambda$ from
the Hamiltonian in favor of a renormalised tunneling amplitude
$t_\alpha \rightarrow t_\alpha e^{-\lambda (\hat{a}-\hat{a}^\dag)}$
and renormalised energy level $\epsilon_0 \rightarrow \epsilon_0'=
\epsilon_0-\lambda^2 \hbar \omega$.
The calculation of the transition matrix elements is then straightforward.
One obtains (see for instance \cite{braig:2003,koch:2005,koch:2006}):
\beqa
    W^{0\rightarrow 1}_{p\rightarrow p',\alpha}
    &=& {\Gamma_\alpha \over \hbar} |M_{pp'}|^2 f(\epsilon_0'+(p'-p)\hbar \omega-\mu_\alpha)
    \label{rate1}
    \\
    W^{1\rightarrow 0}_{p\rightarrow p',\alpha}
    &=& {\Gamma_\alpha \over \hbar} |M_{pp'}|^2
        \left[1-f(\epsilon_0'-(p'-p)\hbar \omega-\mu_\alpha)\right]
    \label{rate2}\,,
\eeqa
where $W^{n\rightarrow n'}_{p\rightarrow p',\alpha}$ is the transition rate
from the occupation state
$n$ to  $n'$ with simultaneous change of the phonon state from
$p$ to $p'$.
In Eqs. \refe{rate1} and  \refe{rate2}
$f(\epsilon)=1/(1+e^{\epsilon/k_B T_e})$ is the Fermi distribution of the electrons
in the metallic leads and $M_{pp'}$ is the matrix element for a phonon transition.
The latter is given by the overlap integral of two harmonic oscillator wavefunctions
spatially displaced by $\Delta x = \lambda \sqrt{2\hbar/m \omega} $,
with $m$ the oscillator mass:
\beq
    |M_{p_1 p_2}|=\lambda^{P-p} e^{-\lambda^2/2} \sqrt{p! \over P!} L_p^{P-p}(\lambda^2)
    \,.
\eeq
The absolute value of the matrix element depends only on $p=\min(p_1,p_2)$
and $P=\max(p_1,p_2)$.
Here $L_p^P$ are the generalized Laguerre polynomials.
We neglect cotunneling events, the reader can find a detailed discussion
of their contribution to the current in Ref.~\cite{koch:2006}.

The mechanical oscillator is coupled to an external environment at a given
temperature $T_m$ that we will assume in general different from
the electronic temperature $T_e$ entering the Fermi distributions.
If the environment correlation time is shorter than all the other time scales
of the problems ($\Gamma/\hbar$ and $1/\omega$) following Ref. \cite{koch:2005}
we can model the effect of the environment with the time scale $\tau$ necessary to reach
the phonon equilibrium state.
The master equation takes then the form:
\beq
    {dP_{np} \over dt }
    =
    -P_{np} W_{np}
    +\sum_{n'p' \alpha }P_{n'p'}W^{n'\rightarrow n}_{p'\rightarrow p,\alpha}
    -{1\over \tau}\left(P_{np}-P^{eq}_{p} \sum_{p'} P_{np'} \right)
    \label{master}
\eeq
where $W_{np}=\sum_{n'p'\alpha}W^{n\rightarrow n'}_{p\rightarrow p', \alpha}$ and
$P_p^{eq}=e^{-p\hbar \omega/k_B T_m}(1-e^{-\hbar \omega/k_B T_m})$
is the canonical equilibrium distribution.
The stationary state of the system can now be obtained by solving
\refE{master} for $dP_{np}/dt=0$.
Electronic transport can drive the oscillator out of equilibrium
and in general the resulting effect can be either to increase or to decrease
the average energy of the oscillator.
In the next Sections we discuss the properties of the
stationary solutions of \refE{master} and the consequences
for the oscillator state.

\section{Cooling by phonon absorption}
\label{sec3}

In this section we provide an analytical solution for the
stationary state of the device and discuss how the bias
condition on the single electron
transistor can influence the thermal state of the
mechanical device inducing either cooling or heating.

Since the stationary solution obtained from
\refE{master} is not in general an equilibrium distribution
strictly speaking we cannot define a temperature for this
state.
Nevertheless one can always define the average phonon energy
$E_{av} \equiv \hbar \omega \sum_{np} p P_{np}$.
In particular, for an equilibrium distribution one has
$ E_{av}= \hbar \omega/(e^{\hbar\omega/k_B T_{m}}-1)$,
thus an effective temperature can always be defined by inverting
it:
\beq
    k_B T_{eff} \equiv \hbar \omega /\ln\left(1+\hbar \omega/E_{av}\right)
    \,.
    \label{effT}
\eeq
In the following to describe the thermal state of the oscillator
we will discuss either the average phonon energy or
the effective temperature of the oscillator defined by \refE{effT}.

\begin{figure}
  \includegraphics[width=9cm,angle=-90]{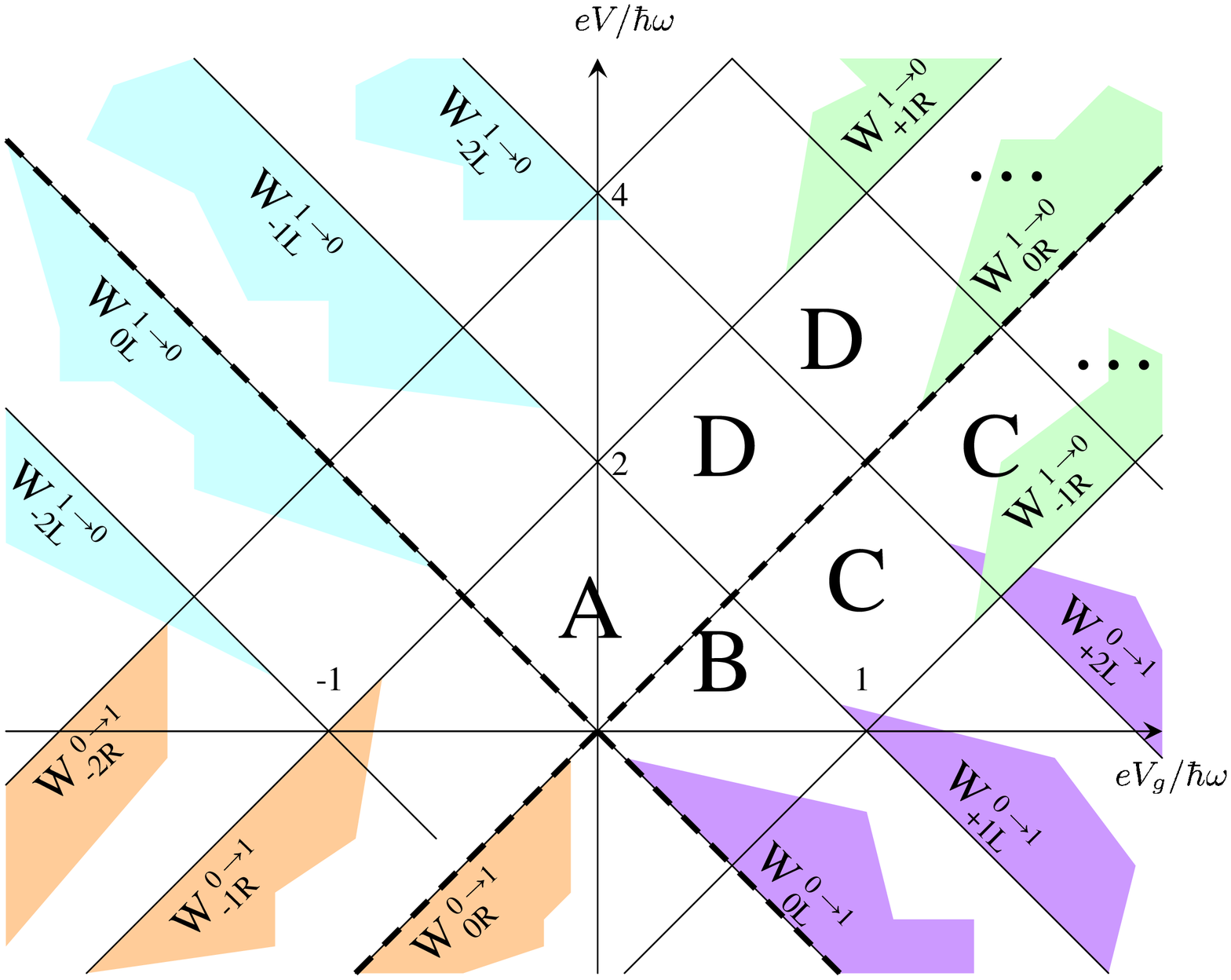}
\caption{(Color online) Regions of non-vanishing rates in the $V_g$-$V$ plane.
The shaded regions with $W^{n\rightarrow n'}_{\Delta p \alpha}$ indicate
the regions where the rate $W^{n\rightarrow n'}_{p\rightarrow p' \alpha}\neq 0$
with $\Delta p = p-p'$.
The dashed line indicates the border of the sequential tunneling transport
regime for the non-interacting case $\lambda=0$.
}
\label{fig1}       
\end{figure}

Let us now discuss the form of the solution qualitatively, and
in particular the contribution of the different tunneling processes.
For simplicity let us assume that the electronic temperature $k_B T_e \ll \hbar \omega\sim k_B T_m$,
then the Fermi functions in \refe{rate1} and \refe{rate2} are, within a good approximation,
step functions.
We define the bias voltage
$eV=\mu_L-\mu_R$ and the gate voltage
$eV_g = (\mu_L+\mu_R)/2 - \epsilon_0'$ in terms of the
chemical potentials of the two leads.
This notation allows to recover the standard description for the
single-electron transistor as a function of $V$ and $V_g$.
For simplicity in the following we concentrate on the $V>0$ and $V_g>0$ case.
(The results are symmetric in both variables if the tunnel junctions are symmetric.)
In this case the electrons are on
average transmitted from the left reservoir to the right reservoir.
Sequential tunneling from the left to the right reservoir
is possible only if some of the $W_{p\rightarrow p',L}^{0\rightarrow 1}$ and $W_{p \rightarrow p',R}^{1\rightarrow 0}$
are non-vanishing at the same time.
For $T_e=0$ the relevant conditions are:
\beq
\left\{
\begin{array}{lcrcl}
    W_{p\rightarrow p',L}^{0\rightarrow 1}\neq 0 & \quad {\rm if} \quad & eV_g-eV/2 &<& -\hbar \omega\Delta p
    \\
    W_{p \rightarrow p',R}^{1\rightarrow 0}\neq 0 & \quad {\rm if}\quad & -eV_g+eV/2 &>& \hbar \omega\Delta p
\end{array}
\right.
\label{diamonds}
    \,.
\eeq
In Fig. \ref{fig1} we show the different regions of non-vanishing
rates, their intersections divide the plane into identical rhombi.
For $\Delta p=(p'-p)=0$  \refE{diamonds} defines
the standard Coulomb diamond conducting region in the $V_g$-$V$ plane,
delimited by the two lines $V=\pm 2V_g$ (dashed in Fig. \ref{fig1}).
The conditions \refe{diamonds} for $\Delta p\neq 0$ divide the
plane in regions where the value of the rate is constant.
The variations are limited to the borders of these regions
and are discontinuous.
%
%
%
%

To draw a simple picture of the mechanism of cooling
we consider in the following the four relevant cases where
the $V$ and $V_g$ are in the regions indicated in the
Fig. \ref{fig1} by A, B, C and D .

Region A is the region of maximal suppression of the current due to
the electron-phonon interaction \cite{braig:2003,koch:2005}.
It is characterized also by a bistability \cite{pistolesi:2007,pistolesi:2008}
since in the limit of large $\lambda$ both the occupied and unoccupied
states are blocked and thus stationary: the final state of the device
depends then on the previous history.
In order to study the stationary state we begin
by assuming that only the probabilities
$P_{n0}$ and $P_{n1}$ are significantly different from zero.
This assumption has to be verified at the end of the calculation
self-consistently and is equivalent to demanding that the
stationary state of the oscillator is essentially the ground state
with a small occupation of the first excited state.
We can then solve the problem with only four states:
$\underline{P}= \{P_{00},P_{01},P_{10},P_{11}\}$.
According to the conditions \refe{diamonds} only few rates are non vanishing in
region A, specifically:
\beqa
    &&  W_{p\rightarrow p,L}^{0\rightarrow 1}=
        W_{p\rightarrow p,R}^{1\rightarrow 0} \equiv W_{Tp}
        \label{traR}
        \\
    &&  W_{1\rightarrow 0,L}^{0\rightarrow 1}=
        W_{1\rightarrow 0,R}^{0\rightarrow 1}=
        W_{1\rightarrow 0,L}^{1\rightarrow 0}=
        W_{1\rightarrow 0,R}^{1\rightarrow 0}
        \equiv W_{C}
        \label{coolingR}
\eeqa where $W_C=\Gamma_0\lambda^2e^{-\lambda^2}/\hbar$ is the
relevant cooling rate ($\Gamma_0 \equiv \Gamma_L=\Gamma_R$).
Note a crucial point, all the rates increasing the number of phonons
(and thus leading to heating) are vanishing.
Only the elastic transport rates \refe{traR}, or the cooling rates
\refe{coolingR} are present, thus one can easily anticipate that
the system will act as a cooler in this regime.
The master equation can be written in matrix form as follows $d\underline{P}/dt = M \underline{P}$
where the matrix $M$ reads
\beq
\left(
    \begin{array}{cccc}
        -W_{T0}-{P_1^{eq}/\tau} & P_0^{eq}/\tau & W_{T0} & 2 W_C \\
        P_1^{eq}/\tau & -W_{T1} -2 W_C-P_0^{eq}/\tau  & 0 &  W_{T1} \\
        W_{T0} & 2 W_C & -W_{T0}-P_1^{eq}/\tau & P_0^{eq}/\tau \\
        0 & W_{T1} & P_1^{eq}/\tau &-W_{T1}-2 W_C -P_0^{eq}/\tau
    \end{array}
\right)
\eeq
and we have introduced the terms due to the coupling
of the oscillator to the external environment.
The stationary solution is given by the eigenvector of vanishing eigenvalue
of the matrix $M$ that reads
\beq
    \underline{P}= \{P_0^{eq}+2\tau W_C, P_1^{eq}, P_0^{eq}+2\tau W_C, P_1^{eq} \}
    \label{solA}
    \,.
\eeq
It gives for the occupation of the excited state $P_q \equiv P_{0q}+P_{1q}$:
\beq
    P_1 = {P_1^{eq}\over1 + 2 W_C \tau}
    \,.
\eeq
This expression has two simple limits:
For $\tau\rightarrow 0$ the environment of the oscillator
dominates and the system remains in thermal equilibrium
with his own bath.
For $\tau\rightarrow \infty$, the system reaches its ground state,
since $P_1\rightarrow 0$.
In general the relevant parameter to determine which of the
two cases is realized is the product $\tau W_C$.
It is maximum for $\lambda=1$ that thus constitutes in this
region the optimal value of the electron-phonon coupling for
cooling efficiency.
Specifically,  for this value of $\lambda$,
$P_1$ reads
$
P_1^{eq}/(1 + 2 \Gamma_0 \tau/e\hbar)
$.
The  condition for efficient cooling
is then simply  $\Gamma_0\tau/\hbar \gg 1$.

To understand better the behavior of the device let us now consider the
three other regions next to the phonon blockade area: regions
B, C and D, as shown in Fig. \ref{fig1}.
In region B the situation is very similar to region A,
the non vanishing rates are
\beqa
    &&  W_{p\rightarrow p,L}^{0\rightarrow 1}=
        W_{p\rightarrow p,R}^{0\rightarrow 1} \equiv W_{Tp}
        \label{elasB}
        \\
    &&  W_{1\rightarrow 0,L}^{0\rightarrow 1}=
        W_{1\rightarrow 0,R}^{0\rightarrow 1}=
        W_{1\rightarrow 0,L}^{1\rightarrow 0}=
        W_{1\rightarrow 0,R}^{1\rightarrow 0}
        \equiv W_{C}\,.
\eeqa
Again we have no heating terms, the only difference
is in the elastic transport terms \refe{elasB}.
We find that the stationary solution for $P_1$ has the
form  \refe{solA} as in the the region A.
The final effective temperature of region B thus coincides
with that of region A.

Let us now consider region C, this is an important case
since one can show that actually all the blocks delimited by
the two lines $eV=2eV_g$ and $eV=2eV_g+2\hbar \omega$ for
$eV_g>-2eV+\hbar \omega$ (indicated as $C$
in Fig. \ref{fig1}) have exactly the same behavior.
The list of non-vanishing rates reads:
\beqa
    &&  W_{p\rightarrow p,L}^{0\rightarrow 1}=
        W_{p\rightarrow p,R}^{0\rightarrow 1} \equiv W_{Tp}
        \label{elasticC}
        \\
    &&  W_{1\rightarrow 0,L}^{0\rightarrow 1}=
        W_{1\rightarrow 0,R}^{0\rightarrow 1}=
        W_{1\rightarrow 0,R}^{1\rightarrow 0}
        \equiv W_{C}
        \label{coolC}
        \\
    &&W_{0\rightarrow 1,L}^{0\rightarrow 1}\equiv W_H = W_C
    \label{heatC}
        \,.
\eeqa
The situation is different from case A and B, since now one
phonon-emission rate is non vanishing [\refE{heatC}].
Still there are three phonon absorption terms \refe{coolC}
and we can expect that the final result will be not
far from the ground state.
Also in this case it is possible to find the stationary solution
analytically, but its form is rather cumbersome, we prefer to present
here only the result for large $\tau$:
\beq
    P_1/P_1^{eq}= {W_C+2W_{T0} \over 2 \tau W_C W_{T0}}
    =
    (\hbar/\tau \Gamma_0) e^{\lambda^2}(2+\lambda^2)/2\lambda^2
    \label{resC}
    \,.
\eeq
We thus find again that for large $\tau$ the single electron transistor
behaves as a good cooler for the oscillator.
It is important to notice that this result holds for
large $eV$, since all the blocks along the threshold of current
behave in the same way.
This holds for any value of $\lambda$, but
the optimal range for cooling is $\lambda \approx 1$.
We can now check the self-consistency of the solution.
The runcation of the space to the first excited state is justified
by the fact that $P_1 \sim \Gamma_0/\tau  $ vanishes for
$\tau\Gamma_0 \rightarrow \infty$.

Let us finally check what happens when the system is biased
in region D, where the sequential transport channel is open.
Only the elastic rates differ from the set of non-vanishing rates of
region C [Eq.: \refe{elasticC} is substituted by the following
Equation] \beqa
    &&  W_{p\rightarrow p,L}^{0\rightarrow 1}=
        W_{p\rightarrow p,R}^{1\rightarrow 0} \equiv W_{Tp}
        \label{elasticD}
        \,.
\eeqa
This difference is very important, since now it is possible
to let an electron out of the central island by elastic tunneling
($W_{q\rightarrow q,R}^{1\rightarrow 0}$) and let an other electron in
by emitting a phonon ($W_{0\rightarrow 1,L}^{0\rightarrow 1}$).
This process heats up the device and is in competition with the
cooling processes.
In region C the only way for an electron to exit the central island is by
absorption of one phonon by the oscillator ($W_{1\rightarrow 0,R}^{1\rightarrow 0}$),
thus also if an electron could enter the device by emitting
one phonon $W_{0\rightarrow 1,L}^{0\rightarrow 1}$ the net balance
of the two processes is always of no energy exchanged.
Since all the other processes imply cooling of the oscillator
the stationary solution is a cold state.
In region $D$ this is no more true, and we can
verify it by finding the stationary state for $\tau=\infty$.
We find that $P_1$ does not vanish in this limit:
\beq
    P_1 = {W_{T0}(W_C+W_{T1}) \over 4 W_{T0} W_{T1} + W_C(3 W_{T0} + W_{T1})}
    \,.
\eeq
Note also that this result for $P_1$ is not the accurate
stationary solution, since the fact that it does not vanish
for large $\tau$ invalidate the truncation of the
space to the the first two phonon states.
To find the true stationary solution we have to include more
states, this can be easily done numerically.
The result will be discussed in the next Section,
but it is clear that the system cannot be cooled at very
low temperature in region D.
It is also clear that going even more deeply inside the
sequential tunneling regime the situation will not
improve, and finally one would find that electronic
transport heats up the oscillator.

\section{Numerical results}
\label{sec4}

In order to confirm the analytical results and to investigate
their limitations in this Section we discuss the numerical
solution of the full master equation \refe{master}.
This can be done by truncating the dimension of the space
to $2 p_M$ states such that the $eV$,$eV_g$,$k_B T_m$, $k_B T_e \ll p_M \hbar \omega$.
In practice in the following $p_M=30$ will be sufficient for our
purposes.

%
%
%
\begin{figure}
  \includegraphics[width=11.cm]{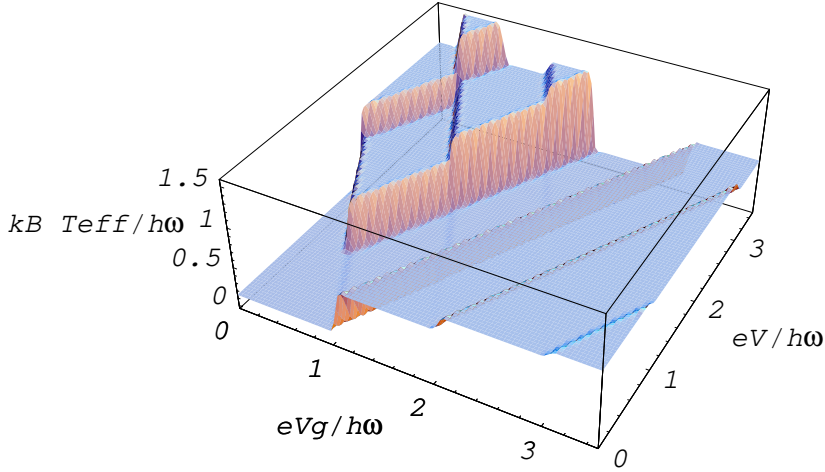}
\caption{(Color online) Effective temperature of the oscillator in the $eV_g$-$eV$ plane.
In this calculation $T_e/\hbar\omega=0.01$, $T_m=\hbar\omega$, $\tau/\Gamma_0=100$, $\lambda=1$, and
$\Gamma_0/\omega=0.1$. We assume symmetric junctions.
In this figure it is easy to recognize the different regions shown
in Fig. \ref{fig1}.}
\label{fig3}       
\end{figure}

We begin by considering an ideal case of very cold electrons $k_B T_e \ll \hbar \omega$,
and long relaxation rate of the phonons $\tau \Gamma_0/\hbar =100$, with
$\Gamma_0/\hbar \omega=0.1$ and $\lambda=1$.
The resulting effective temperature of the oscillator is shown
as a function of the bias and gate voltage in Fig. \ref{fig3}.
The plot shows how important are the bias condition,
since the effective temperature dependence clearly allows to
recognize the different regions we discussed in Section \ref{sec3},
since the temperature changes abruptly at their borders.
The results confirm the analysis of the previous Section,
regions A, B, and the stripe of C regions correspond to a stationary
solution with very low effective temperature (cfr. Fig. \ref{fig3}).
The effective temperature is much smaller than the mechanical reservoir temperature.
In particular the lowest  temperature is obtained for any of the
$C$ regions, regardless of the value of the bias voltage.
We also can see that the temperature is higher in the D rhombi
for the reasons explained in Section \ref{sec3}.
%

%
%
%
\begin{figure}
\includegraphics[width=8.cm]{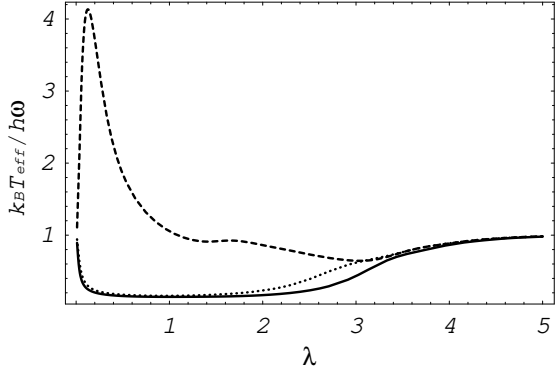}
\caption{Effective temperature as a function of the coupling
constant $\lambda$ for the same parameters of Fig. \ref{fig3}
for different bias points.
Specifically the full, dotted, and dashed lines correspond to
the bias points $(eV_g,eV)/\hbar\omega=(0,0.5)$, $(0,2.5)$,
and $(1,0.5)$, respectively.
}
\label{fig4}       
\end{figure}

In Fig. \ref{fig4} we plot the coupling dependence of the
effective temperature for the same parameters of the
Fig. \ref{fig3} at different bias points as indicated
in the caption.
The full and dotted line correspond to the case A, and C,
respectively. We see that the optimal cooling is attained
for $\lambda=1$, as found in the analytic calculation.
The dashed line shows the behavior in the middle of the
Coulomb diamond, and for bias voltage larger that
$2\hbar \omega$, that constitutes the inelastic threshold
in this case.
We see that an increase of the coupling leads to a strong
heating of the device, which is expected since the electrons
have enough energy to release one phonon in the oscillator
every time they traverse the device.
For very large coupling the effect is suppressed by the
reduction of all the rates due to the
exponential vanishing of the overlap of the
harmonic oscillator wavefunctions.

%

%
%
%
\begin{figure}
  \includegraphics[width=11.cm]{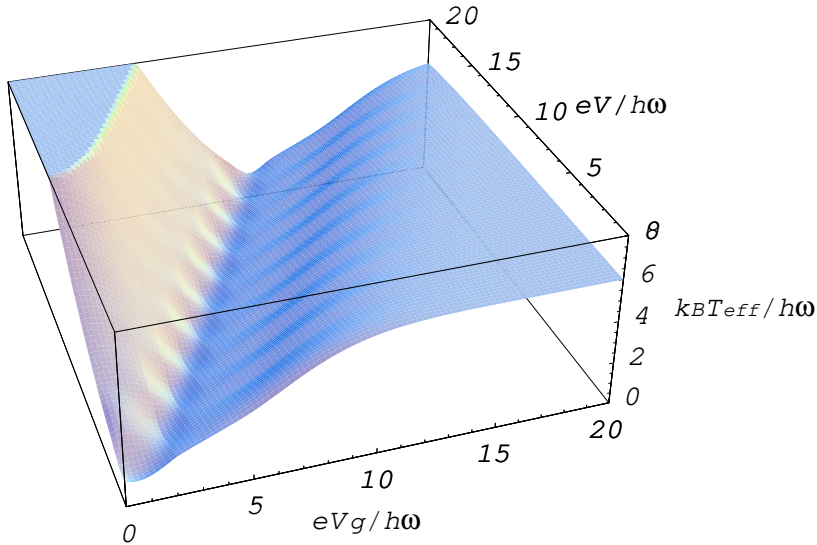}
\caption{(Color online) Effective temperature for
$T_e/\hbar\omega=0.1$, $T_m/\hbar\omega=5$, $\tau/\Gamma_0=100$, $\lambda=1$, and
$\Gamma_0/\omega=0.1$.
Compared  to Fig. \ref{fig3} we plot here a much larger region of the
plane, since the temperature is higher.
}
\label{fig5}       
\end{figure}

Finally let us consider the case of $T_m=5 \hbar \omega$ and
still cold electrons $T_e=0.1 \hbar \omega$.
The effective temperature for this case (the other parameters
are the same of Fig. \ref{fig3}) is shown in Fig. \ref{fig5}.
The main feature is clearly the presence of a deep trench
at the border of the conducting diamond.
Again this is in agrement with the analytical results, that
shows that $T_{eff}$ should not depend on $eV$ along that
line.
We also see that in the conducting region for large enough
voltages the current through the device heats up the
oscillator, as is usually the case.
On the other side, too far from the conducting region
the temperature is not modified since transport is
completely suppressed.

As a final check we also verified that when the electronic temperature
$T_e$ coincides with $T_m$, cooling is not possible.
There is always a dependence of the effective temperature
of the stationary state on the bias condition, and in some case
we find a very small reduction of the temperature of the
oscillator, but clearly in this regime the device could not
be used as a cooler for the oscillator.

\section{Conclusions}
\label{sec5}

We have shown that a molecular (single electronic level) single-electron
transistor coupled to an oscillator can act as a cooler for the
oscillator if the electronic temperature is lower than the
temperature of the environment to which the oscillator is
coupled.
The heat transport mechanism is not trivial since it involves
the phonon absorption and emission, that can be finely tuned
by the bias condition of the single electron transistor.
We find that optimal cooling is achieved by biasing the device
near the border of the sequential tunneling conducting region of
the $V_g$-$V$ plane.
By using for instance normal-metal-superconductor tunnel junctions
electronic cooler, it is possible to generate cold sources of electrons and
then to exploit them to cool nano-electromechanical oscillators.

\begin{acknowledgements}
I acknowledge a stimulating discussion on this problem
with A. Armour and M. Houzet.
This work has been supported by the French {\em Agence Nationale
de la Recherche} under contract ANR-06-JCJC-036 NEMESIS.
I thank A. Buzdin and his group for hospitality at the { \em Centre de Physique
Moleculaire Optique et Hertzienne} of Bordeaux (France) where part of this work has
been completed.
\end{acknowledgements}

\bibliographystyle{spphys}       


\bibliography{biblioNEMS,biblio}

\end{document}